# Continuum limit of field theories regularized on a random lattice [*]

B. Allés[a][†], M. Beccaria[a], L. Del Debbio[a] and R. Del Real[b]

[a]Dipartimento di Fisica, Università di Pisa and INFN,
Piazza Torricelli 2, 56126-Pisa, Italy

[b]Departamento de Física Teórica y del Cosmos, Facultad de Ciencias,
Universidad de Granada, 18071-Granada, Spain

The continuum limit and scaling properties of an asymptotically free field theory regularized on a random lattice are compared with those on a regular square lattice. We work on random lattices parametrized by a degree of "randomness" $\kappa$. We show that the continuum limit exists and different $\kappa$ are related by a finite renormalization.

## 1. INTRODUCTION

The lattice regularization was introduced to study the non-perturbative aspects of field theories [1]. To extract continuum physics information from the numerical Monte Carlo, the simulations must be performed in a region where the correlation length $\xi$, the lattice spacing $a$ and the lattice size $L$ satisfy $1 \ll \xi/a \ll L$. How strong these inequalities are depends on the theory we are studying and the particular lattice used. The random lattice was put forward to weaken the above inequalities [2]. Indeed, the isotropic properties of the random lattice should expedite the continuum limit.

However, the random lattice has been little exploited for numerical studies of field theories. On the other hand, no much work has been done to check whether its continuum limit is the same than that of the usual regular square lattice. In this work we have considered the $O(3)$ non-linear $\sigma$-model in two dimensions in order to address the above questions.

To construct the random lattice, we followed the procedure of T. D. Lee et al. [2,3]. The only new ingredient is the introduction of a degree of "randomness" $\kappa$ [4]. The sites of the lattice are the centers of hard spheres, the radius of which is $a/(2\kappa)$. These hard spheres are randomly located and their relative distance follows the Poisson distribution. At small values of $\kappa$ the lattice turns out to be locally less random, (i.e.: link lengths, plaquette areas, coordination numbers, etc. have smaller variance)[4].

## 2. THE $O(3)$ $\sigma$-MODEL ON THE RANDOM LATTICE

Following [5], the action of the model on a random lattice can be written as

$$\mathcal{S}^{\mathrm{rand}} = \frac{1}{4g} \sum_{i,j} \lambda_{ij} \left(\vec{\phi}_i - \vec{\phi}_j\right)^2. \qquad (1)$$

In this expression, $g$ is the coupling constant. The matrix $\lambda_{ij}$ is the weight that allows Eq.(1) to have the correct naïve continuum limit. It is defined as the ratio between the lengths of the dual link and the link joining the sites $i$ and $j$. $\vec{\phi}_i$ is the value of the field $\vec{\phi}$ at the site $i$. In a recent paper, [4], it has been shown that this model on a random lattice is asymptotically free and the renormalization group invariant scale $\Lambda$ depends on $\kappa$, the "randomness" degree parameter.

Beside these perturbative results, an explicit numerical simulation gives us an insight about non-perturbative features such as the mass gap and the topological content, which can be compared to the continuum results [6–8].

To measure the mass gap, we computed the 2-point correlation function at zero spatial momentum and fitted the result with an hyperbolic cosinus.

---

As a definition of the topological charge operator on the random lattice we propose

$$Q_i = \frac{1}{32\pi}\frac{1}{\omega_i}\epsilon_{abc}\sum_{j,k} \phi_i^a\,\phi_j^b\,\phi_k^c\,\lambda_{ij}\,\lambda_{ik}\,\sigma_{ijk}, \quad (2)$$

where $Q_i$ is the topological charge density at the site $i$, $\omega_i$ the volume of the Voronoi cell around this site [2] and $\sigma_{ijk}$ is defined as

$$\sigma_{ijk} \equiv \left[\left(\vec{l}_{ij}+\vec{d}_{ij}\right)\times\left(\vec{l}_{ik}+\vec{d}_{ik}\right)\right]\cdot\hat{z}. \quad (3)$$

In this expression, $\vec{l}_{ij}$ is the link vector joining the sites $i$ and $j$ and $\vec{d}_{ij}$ is the vector joining the center of the above link with the center of the associated dual link [5]. Finally, $\hat{z}$ is the unit vector orthonormal to the plane of the lattice, oriented as $\hat{z} = \hat{x}\times\hat{y}$. One can prove that $Q_i$ has the correct *naïve* continuum limit. The total topological charge $Q$ of a configuration is obtained simply by $Q = \sum_i Q_i$. On a regular lattice, our topological charge coincides with the definition of reference [9].

## 3. NUMERICAL RESULTS

In the Monte Carlo simulation we used the Wolff updating algorithm [10]. Depending on $\beta = 1/g$, $2 \div 10 \times 10^4$ initial sweeps are sufficient to thermalize the configuration. In order to compare the performance on each lattice, we repeated the same measurement on a regular square lattice and on a $\kappa = 100$ and $\kappa = 1.3$ random lattices. These values of $\kappa$ correspond to rather opposite levels of "randomness", the first being completely irregular and the second almost regular. We did not average the result among several random lattices since we have used a large enough number of sites. On the regular lattice we used the standard action, and the action of Eq.(1) on the random lattice. Here, we report the results obtained with lattices ranging from $200^2$ to $400^2$ sites.

### 3.1. Mass measurement

For every value of $\beta$, 12000 updatings were made. On each of them the Wolff's improved estimator was used to compute the 2-point correlation function [11]. We checked that the correct continuum limit is reached by fitting the exponent $\eta$ in the 2-loop renormalization group prediction

$$\text{Monte Carlo data} = \frac{m}{\Lambda}\,\zeta\beta\,\mathrm{e}^{-\eta\beta}. \quad (4)$$

The universal beta function dictates $\eta = \zeta = 2\pi$. We found that $\eta$ is equal to $2\pi$ within 1% on regular and random lattices independently of the $\kappa$ used. However, we could not measure $\zeta$ which is very sensitive to the non-universal terms. Correcting the data for finite size effects as shown in [12,13] did not change this situation.

In figure 1 we report the ratio between $\Lambda$ parameters for both $\kappa = 100$ and $\kappa = 1.3$. The average values are 1.29(8) and 0.86(6) respectively, and do not change when increasing the lattice size. The theoretical calculated values are 1.8(2) and 0.92(2) respectively [4]. We think that the disagreement between the calculated and measured ratios for $\kappa = 100$ may be due to large non-universal terms in the beta function. On the other hand, we see a rather good agreement in this ratio for $\kappa = 1.3$, suggesting a $\kappa$ dependence of the non-universal terms in the beta function. Preliminary results for the $O(4)$ and $O(8)$ $\sigma$-models show a much better agreement with the perturbation theory predictions [14].

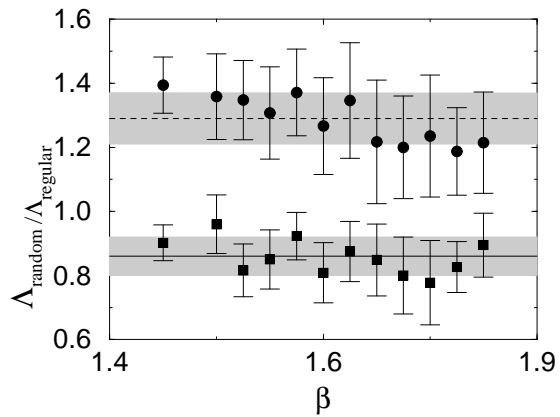

Figure 1. Ratio of $\Lambda$ parameters for $\kappa = 100$ (dashed line and circles) and $\kappa = 1.3$ (solid line and squares) calculated on a $200^2$ lattice.

## 3.2. Topological Susceptibility

We also measured the topological susceptibility, defined as

$$\chi = \frac{1}{V} \langle Q^2 \rangle, \qquad (5)$$

where $V$ is the lattice volume, using the cooling method adapted to the random lattice action. The topological charge was measured after 30 cooling steps for each uncorrelated configuration [9]. It is interesting to notice that the same clustering towards integer values observed after the cooling on regular lattices holds here too. We checked that this measurement was unchanged within errors after 40 and 50 cooling steps. We also verified that the result for the topological susceptibility is insensitive to the rounding of the previous value of the topological charge to the nearest integer.

We measured the topological charge on 1000 uncorrelated configurations separated by 500 updatings and checked that there was no autocorrelation at all. The Monte Carlo data provided a value for the $\eta$ parameter in the exponent of the renormalization group prediction a $\sim 20\%$ away from the perturbative value $2\pi$ for both random and regular lattices. This discrepancy diminishes when using larger lattices.

An analysis, analogous to that of figure 1, applied on the topological susceptibility data, produces similar results: 1.37(3) and 0.94(2) for $\kappa = 100$ and $\kappa = 1.3$ respectively.

## 4. CONCLUSIONS

By varying the $\kappa$ parameter, we can construct random lattices with different levels of "randomness". We checked that they correspond to different regularizations so that any average among random lattices must take into account this fact. Our Monte Carlo data support the scenario of a common continuum limit as was put forward in [4]. The theory regularized on a random lattice presents a mass gap which scales as it should. Moreover, the Monte Carlo data indicate that there is a topological content as well.

We remark that the non-universal terms in the beta function are large enough to prevent the onset of asymptotic scaling (we could not determine the coefficient $\zeta$). In that sense, the random lattice, compared with the regular lattice, does not improve the asymptotic scaling.

A similar analysis for QCD is in progress. There, the $\Lambda_{random}$ is much less than $\Lambda_{regular}$ [15]. This fact shifts the scaling window to larger betas in a much more effective way than for the $O(3)$ $\sigma$-model, possibly allowing a better control on the perturbative expansions that mask the non-perturbative signal.

We wish to thank Ettore Vicari for useful conversations and Andrea Viceré for help in the computer simulations. We also acknowledge financial support from INFN (Italy).

## REFERENCES

1. K. Wilson, Phys. Rev. **D10** (1974) 2455.
2. N. H. Christ, R. Friedberg and T. D. Lee, Nucl. Phys. **B202** (1982) 89.
3. R. Friedberg and H. C. Ren, Nucl. Phys. **B235** [**FS11**] (1984) 310.
4. B. Allés, Pisa preprint "The random lattice as a regularization scheme", hep-lat/9405008.
5. N. H. Christ, R. Friedberg and T. D. Lee, Nucl. Phys. **B210** (1982) 337.
6. A. A. Belavin and A. M. Polyakov, JETP Letters **22** (1975) 245.
7. P. Hasenfratz, M. Maggiore and F. Niedermayer, Phys. Lett. **B245** (1990) 522.
8. E. Brézin and J. Zinn-Justin, Phys. Rev. **B14** (1976) 3110.
9. A. Di Giacomo, F. Farchioni, A. Papa and E. Vicari, Phys. Rev. **D46** (1992) 4630.
10. U. Wolff, Phys. Rev. Lett. **62** (1989) 361.
11. U. Wolff, Nucl. Phys. **B334** (1990) 581.
12. I. Bender and W. Wetzel, Nucl. Phys. **B269** (1986) 389.
13. M. Lüscher in "Progress in gauge field theory" (Cargèse 1983), G. 't Hooft et al. (eds.), Plenum, New York (1984).
14. Work in preparation.
15. Z. Qiu, H. C. Ren, X. Q. Wang, Z. X. Yang and E. P. Zhao, Phys. Lett. **B203** (1988) 292.